\documentclass[iop]{emulateapj}

\slugcomment{Accepted for Publication in the Astrophysical Journal, 17 May 2013}

\shorttitle{Dusty Filaments in Hercules A}
\shortauthors{O'Dea et al.}

\usepackage{epstopdf}
\usepackage{apjfonts}
\def\aj{AJ}%
\def\araa{ARA\&A}%
\def\apj{ApJ}%
\def\apjl{ApJ}%
\def\apjs{ApJS}%
\def\ao{Appl.~Opt.}%
\def\apss{Ap\&SS}%
\def\aap{A\&A}%
\def\aaps{A\&AS}%
\def\mnras{MNRAS}%
\def\pasp{PASP}%
\def\nat{Nature}%
\def\physrep{Phys.~Rep.}%
\def\lae{\mathrel{<\kern-
1.0em\lower0.9ex\hbox{$\sim$}}} 
\def\gae{\mathrel{>\kern-
1.0em\lower0.9ex\hbox{$\sim$}}}
 \def\deg{^{\circ}}

 \def\mone{$^{-1}$} 
\def\mtwo{$^{-2}$}

\begin{document}


\title{{\it Hubble Space Telescope} Observations of Dusty Filaments in Hercules A:  \\Evidence for Entrainment}


\author{Christopher P.~O'Dea}
\author{Stefi~A.~Baum}
\affil{School of Physics and Astronomy, Laboratory for Multiwavelength Astrophysics, and Chester F.~Carlson Center 
for Imaging Science, Rochester Institute of Technology, 84 Lomb Memorial Dr., Rochester, NY 14623, USA}

\author{Grant R.~Tremblay}
\affil{European Southern Observatory, Karl-Schwarzschild-Str. 2, 85748 
Garching bei M\"{u}nchen, Germany}

\author{Preeti Kharb}
\affil{Indian Institute of Astrophysics, II Block, Koramangala, Bangalore 560034, India}

\and

\author{William~D.~Cotton}
\author{Rick~A.~Perley}
\affil{National Radio Astronomy Observatory, 520 Edgemont Road, Charlottesville,
VA 22903, and P.O. Box 0, Socorro, NM 87801, USA}





\begin{abstract}
We present U, V, and I-band images of the host galaxy of Hercules A (3C 348) obtained with HST/WFC3/UVIS. We find a network of dusty filaments which are more complex and extended than seen in earlier HST observations. The filaments are associated with a faint blue continuum light (possibly from young stars) and faint H$\alpha$ emission. It seems likely that the cold gas and dust has been stripped from a companion galaxy now seen as a secondary nucleus. There are dusty filaments aligned with the base of the jets on both eastern and western sides of the galaxy. The morphology of the filaments is different on the two sides - the western filaments are fairly straight, while the eastern filaments are mainly in two loop-like structures. We suggest that despite the difference in morphologies, both sets of filaments have been entrained in a slow moving boundary layer outside the relativistic flow. As suggested by \citet{fabian08}, magnetic fields in the filaments may stabilize them against disruption.  We consider a speculative scenario to explain the relation between the radio source and the shock and cavities in the hot ICM seen in the Chandra data \citep{nulsen05}. We suggest the radio source originally ($\sim 60$ Myr ago) propagated along a position angle of $\sim 35\deg$ where it created the shock and cavities.  The radio source axis changed to its current orientation ($\sim 100\deg$) possibly due to a supermassive black hole merger and began its current epoch of activity about 20 Myr ago.  
\end{abstract}





\section{Introduction}

\object{Hercules A} (3C 348) has the  fourth highest flux density at 178 MHz.  The two radio jets, extended over $\sim 400$ kpc, have remarkably different morphology \citep{dreher84,gizani03,cotton13}
 with the Eastern side showing a twisting jet, and the Western side showing
a series of ring-like features. There is no consensus explanation for the difference
in radio morphology \citep{mason88, meier91, sadun02, saxton02, nakamura08}.

The host galaxy is the dominant cD in a poor cluster at redshift z=0.154 \citep{greenstein62,owen89, allington93, zirbel96}
 and contains a  secondary nucleus  \citep[e.g.][]{smith89,sadun93,baum96, ramos11}.
Although poor optically, the cluster is luminous in X-rays
and contains a cooling flow \citep{gizani04,nulsen05}.
There are two $\sim 40$ kpc-scale cavities in the X-ray emitting gas that are nearly perpendicular to  the radio axis \citep{nulsen05}.
There is a strong shock in the hot gas indicating the release by the radio source of $\sim 10^{61}$ ergs of mechanical energy \citep{nulsen05}.

There are dusty filaments in the host galaxy seen in HST \citep{baum96, dekoff96}  
and ground-based \citep{ramos11} images.  In this paper we present deeper HST/WFC3 observations,
examine the nature of  the filaments in more detail and discuss their relation to the radio source. 
We adopt a cosmology with H$_o = 71$ km s\mone Mpc\mone, $\Omega_M = 0.27$ and
$\Omega_\Lambda = 0.73$ which gives a scale of 2.64 kpc/arcsec at the distance of
Hercules A (D$_L = 726$ Mpc).

\begin{figure*}
\begin{center}
\includegraphics[scale=0.5]{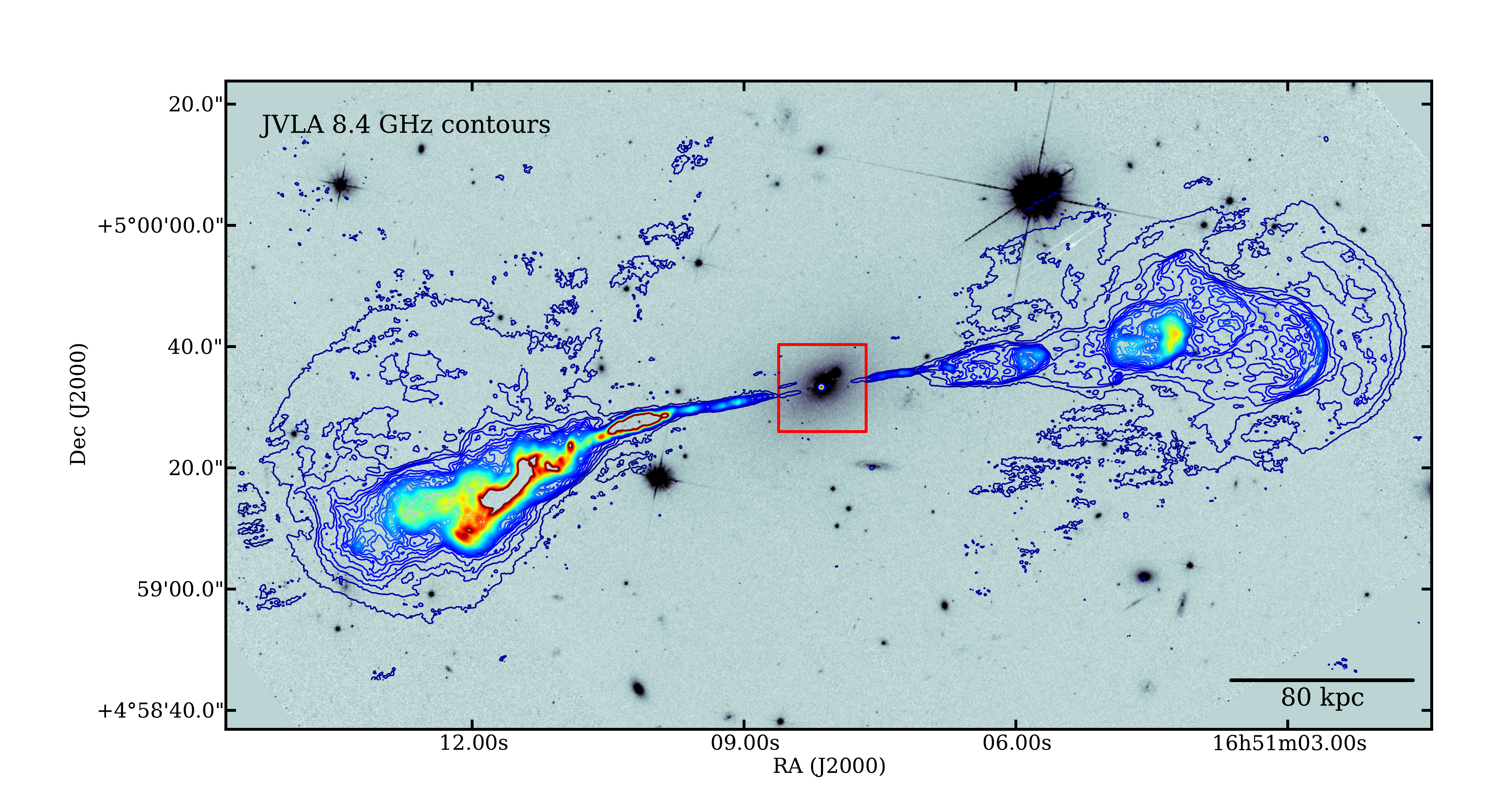}
\end{center}
\caption{Hercules A.   HST/WFC3/UVIS V-band (F606W) image with contours of the 8.4 GHz JVLA image \citep{cotton13}  superposed 
 showing the visible extent of the radio source. The colors of the JVLA radio contours range from blue to red with increasing 8.4 GHz intensity. The red box shows the approximate field of view for the subsequent figures. \label{fig_JVLA}} 
\end{figure*}

\section{HST Observations\label{sec_obs}}

The HST observations were obtained on October 8, 2012 during a 3-orbit visit. We obtained U, V, and I images with the
WFC3/UVIS camera.   The observation parameters are given in Table 1. The observations were dithered to allow removal of cosmic rays and bad pixels and to cover the WFC3 chip gap. The HST calibration pipeline was applied to the raw data.   Figure \ref{fig_JVLA} shows theV-band image and the full extent of the deep  Jansky Very Large Array   (JVLA) image from \citet{cotton13}. The   U, V, and I images are shown in Figures  \ref{fig_U}, \ref{fig_V},  and \ref{fig_I}.

The optical emission lines are faint and low excitation \citep{tadhunter93, buttiglione09}. H$\alpha$+[NII] and [OII] are the  brightest emission lines in the optical spectrum  \citep{buttiglione09}. The U-band (F336W) filter  lies blue-ward of [OII] and the V-band (F606W) filter sits between H$\alpha$+[NII] and [OII].  Thus the U and V  filters are dominated by the continuum in the host galaxy.  There will be some contamination of the I-band (F814W) filter by  H$\alpha$+[NII]. Our rough estimate suggests that contamination from  H$\alpha$+[NII] to the total flux is less than 20\%, though this will be spatially dependent.

Galactic (Milky Way) extinction was corrected using an E(B-V) = 0.25. Internal extinction was corrected using an E(B-V) = 0.34. This was calculated from the dereddened (H$\alpha$/H$\beta$) Balmer decrement ratio from the optical spectroscopy of \citet{buttiglione09}, using 

$E\left(B-V\right)_{H\alpha / H\beta} = \frac{2.5 \times \log \left( 2.86 / R_{\mathrm{obs}}\right)}{k\left(\lambda_\alpha\right) - k\left(\lambda_\beta\right)}$

where R$_\mathrm{obs}$  is the Balmer decrement ($\sim 4.0$) from
\citet{buttiglione09}, and K($\alpha$) = 2.54, K($\beta$)=3.61 for the
R$_\mathrm{V}\ $= 3.1 extinction law of \citet{cardelli89}. The E(B-V) derived from the Balmer Decrement was then converted into extinctions at U- and I-band to correct our U- and I-band photometry for internal extinction, using Table 3 of \citet{cardelli89}. Final extinctions were A$_\mathrm{Uband-internal}$ = 1.654 and A$_\mathrm{Iband-internal}$ = 0.504.

\begin{deluxetable}{lrr}
\tablecaption{HST Observations \label{tbl-HST}}
\tablewidth{0pt}
\tablehead{
\colhead{Band} & \colhead{Filter }  & \colhead{Int. Time (sec)}
}
\startdata
U-Band & F336W & 3656 \\
V-band & F606W & 1760 \\
I-band & F814W & 1928 \\
\enddata
\tablecomments{The HST observations were taken under Program ID 13065 (PI. S. Baum). The filter name in the second column contains the approximate central wavelength in nanometers.}
\end{deluxetable}

\section{Results}

\subsection{The Nucleus}\label{sec_nucleus}

\begin{figure}
\begin{center}
\includegraphics[scale=0.37]{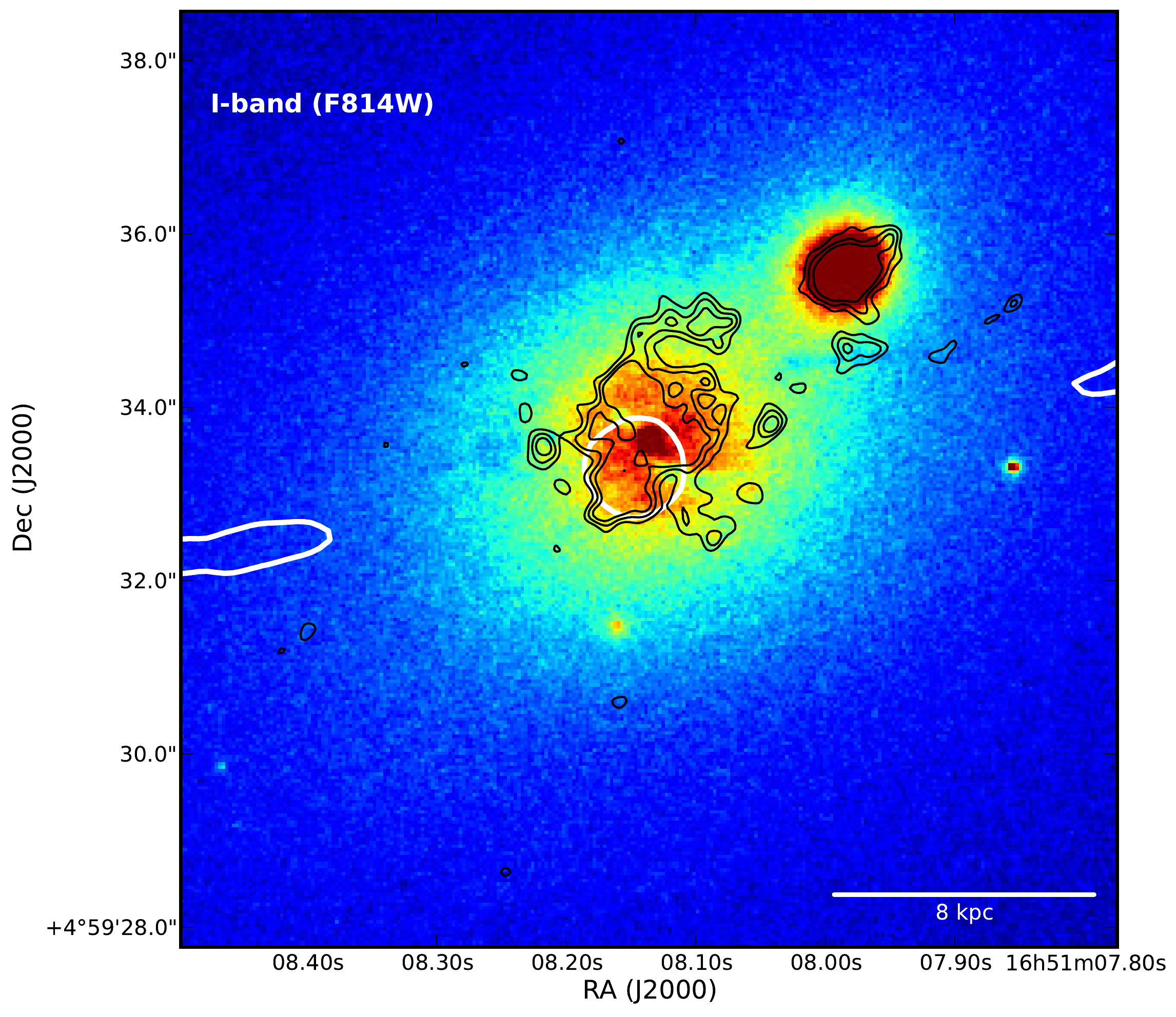}
\end{center}
\vspace*{-5mm}
\caption{Hercules A.  
 HST/WFC3/UVIS I-band (F814W) image.  We detect the host galaxy nucleus at RA = 16h 51m 08.14s Dec = +04$^o$  59$\arcmin$ $33\farcs 6$ in the HST WCS.  
The nucleus seems likely to be dominated by  H$\alpha$+[NII] (\S \ref{sec_obs},\ref{sec_nucleus}). The secondary nucleus is 4\arcsec\ to the north-west of the nucleus of the host galaxy. Low level radio contours of the JVLA image are superposed in white. U-band (F336W) contours are superposed in black.  \label{fig_I}}
\end{figure}

In our  I-band image (Figure \ref{fig_I}) we see a compact nucleus, which is not seen in our V-band image.  The detection in I-band  allows registration of the HST images  with the JVLA image.  The I-band flux of the nucleus corrected for Galactic extinction is $3.34 \pm 0.77 \times 10^{-18}$ ergs s$^{-1}$ cm$^{-2}$ $\AA^{-1}$.
 In contrast, the nucleus was not  detected in the NICMOS H-band \citep{baldi10}  with an upper limit on the monochromatic power of $2.2 \times 10^{28}$ ergs s\mone\ Hz\mone\ which corresponds to a limit on flux of $4.1\times 10^{-18}$ s$^{-1}$ cm$^{-2}$ $\AA^{-1}$.  We convolved the optical spectrum from \citet{buttiglione09} with the WFC3 U, V, and I bandpasses. The continuum in the spectrum is fairly flat between V and I; while the H$\alpha$+[NII] lines (redshifted to $\sim 7575$\AA\ and with equivalent width $\sim 50$\AA)  make a strong contribution to the I band flux. Thus, the detection of the nucleus in I band only seems likely to be due to the H$\alpha$+[NII] lines. 
Our images also clearly show the secondary nucleus previously detected by e.g., \citet{smith89}and \citet{sadun93}. We are not aware of a redshift for the secondary nucleus.

\begin{figure}
\begin{center}
\includegraphics[scale=0.37]{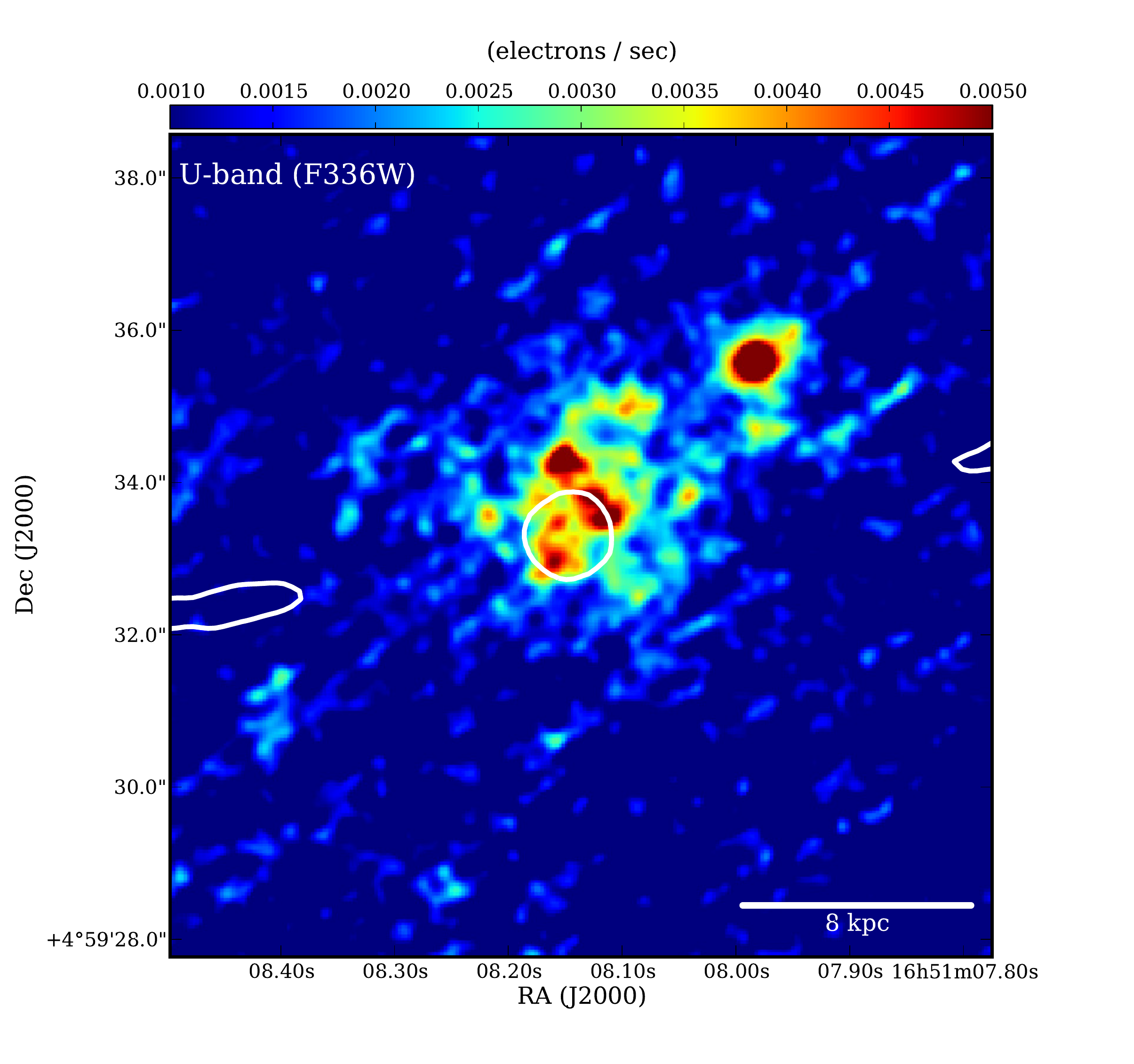}
\end{center}
\vspace*{-5mm}
\caption{Hercules A.  HST/WFC3/UVIS U-band (F336W)  image smoothed slightly with low level radio contours of the JVLA image superposed. This image is not sensitive to the older stellar population. The compact feature 4\arcsec\  to the north-west is the secondary nucleus   at RA = 16h 51m 07.98s Dec = +04$^o$  59$\arcmin$ $35\farcs 6$ in the HST WCS.   \label{fig_U}}
\end{figure}

\begin{deluxetable}{lr}
\tablecaption{Position Angles of Key Components\label{tbl-PA}}
\tablewidth{0pt}
\tablehead{
\colhead{Feature} & \colhead{PA (deg) } 
}
\startdata
Inner radio jets & 100 \\
cD Major Axis  & 120  \\
Secondary Nucleus & 135 \\
Overall Dust Filaments & $\sim 100$ \\
Inner eastern filament & 87 \\
Outer eastern filament & 111 \\
X-ray Cavities & $\sim 35$ \\
\enddata
\tablecomments{Position Angle is measured East from North. }
\end{deluxetable}

\subsection{The Dusty Filaments}

\begin{figure}
\begin{center}
\includegraphics[scale=0.37]{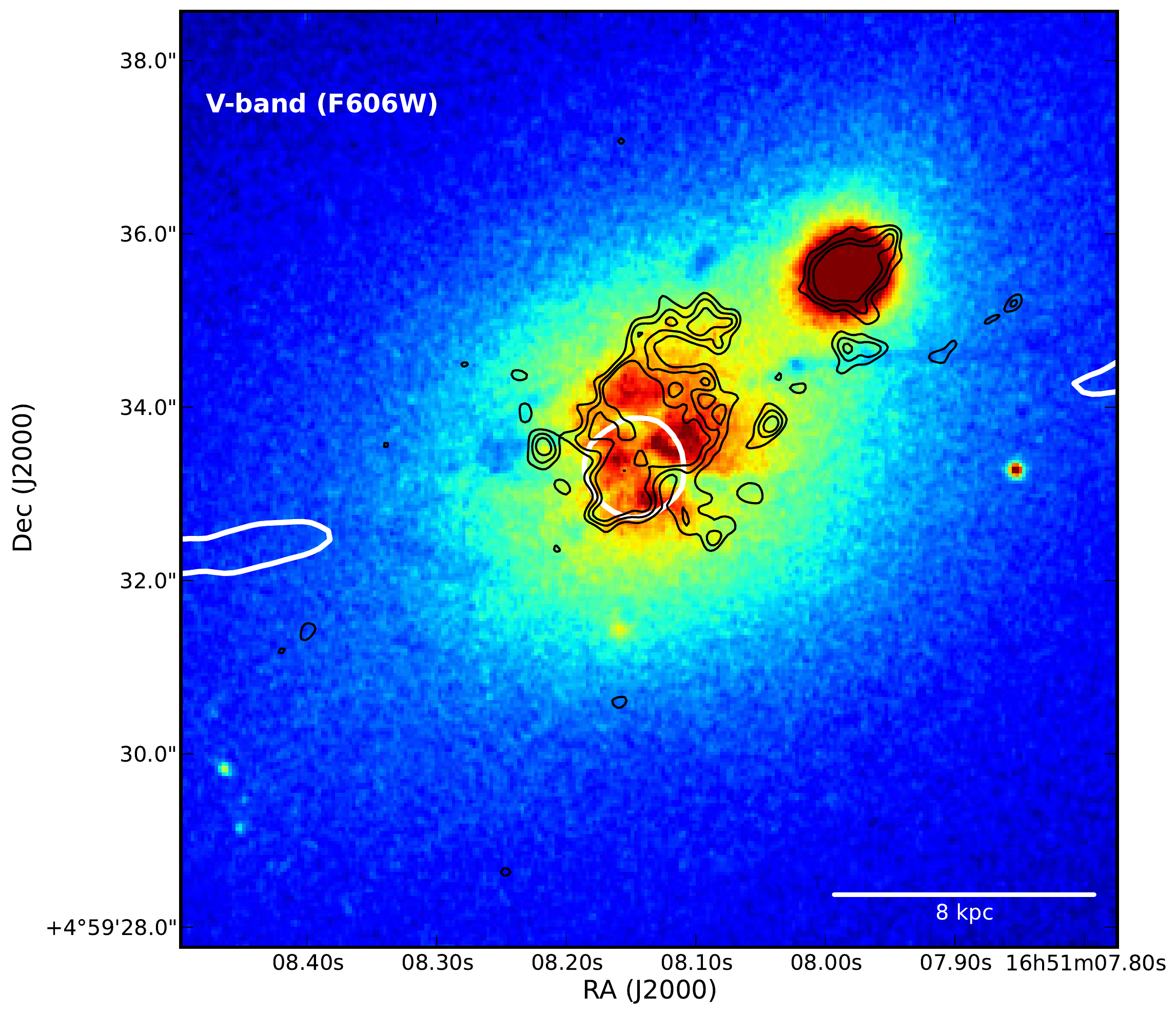}
\end{center}
\vspace*{-5mm}
\caption{Hercules A.  HST/WFC3/UVIS V-band (F606W)  image with low level radio contours of the JVLA image superposed in white. U-band (F336W) contours are superposed in black. \label{fig_V} } 
\end{figure}

The dusty filaments can be seen in   Figures \ref{fig_V} and \ref{fig_dashed}  in the regular and unsharp-masked V-band images. 
The deeper HST images presented here show a network of dusty filaments which are more extended and more
complex than the structures seen in the previous  HST data \citep{baum96, dekoff96}. The largest angular
 size of the filament  web is $\sim 10\arcsec$  ($\sim 26$ kpc).
Position angles of the cD galaxy, secondary nucleus, and inner radio jets  are given in Table 2.
The overall position angle of the filaments is oriented along the radio jet
at PA $\sim 100\deg$, though there is considerable sub structure.
Of course, the morphology of the filaments differs on the two sides of the source. On the western side, the filaments
tend to be relatively straight. Two filaments seem to bracket the location of the western radio jet (projected in towards
the nucleus), while a third extends northwest up to the east of the secondary nucleus. There is a blob of obscuration
just to the east of the secondary nucleus.  On the eastern side the filaments are in
 two curved structures resembling parts of two loops - one extending $\sim 2\arcsec$ and the other $\sim 4\arcsec$ 
from the nucleus.

Figure \ref{fig_RGB} shows a RGB image based on the I, V, and U-band images. We see that there are faint patches of blue light
centered on the galaxy nucleus with a total extent of $\sim 3\arcsec$ (8 kpc) with some additional small  patches further out. Some of the blue patches are spatially adjacent to the dusty filaments.

\section{Discussion}

\subsection{Origin of the dusty filaments \label{sec_origin}}

Hercules A is in a cooling flow cluster \citep{nulsen05}  with an estimated mass accretion rate of 90 M$_\odot$ yr\mone\ (\S \ref{sec_SFR}).
 Cooling flows are thought to deposit cold gas in the central dominant
galaxies though at rates of  about $\sim 10\%$  of those in  the original cooling flow models \citep[e.g.,][]{edge01, peterson06, odea08}. 
 Thus, it is possible that the filaments are the result of gas condensing out of  the cooling flow. 

\begin{figure*}
\begin{center}
\includegraphics[scale=0.65]{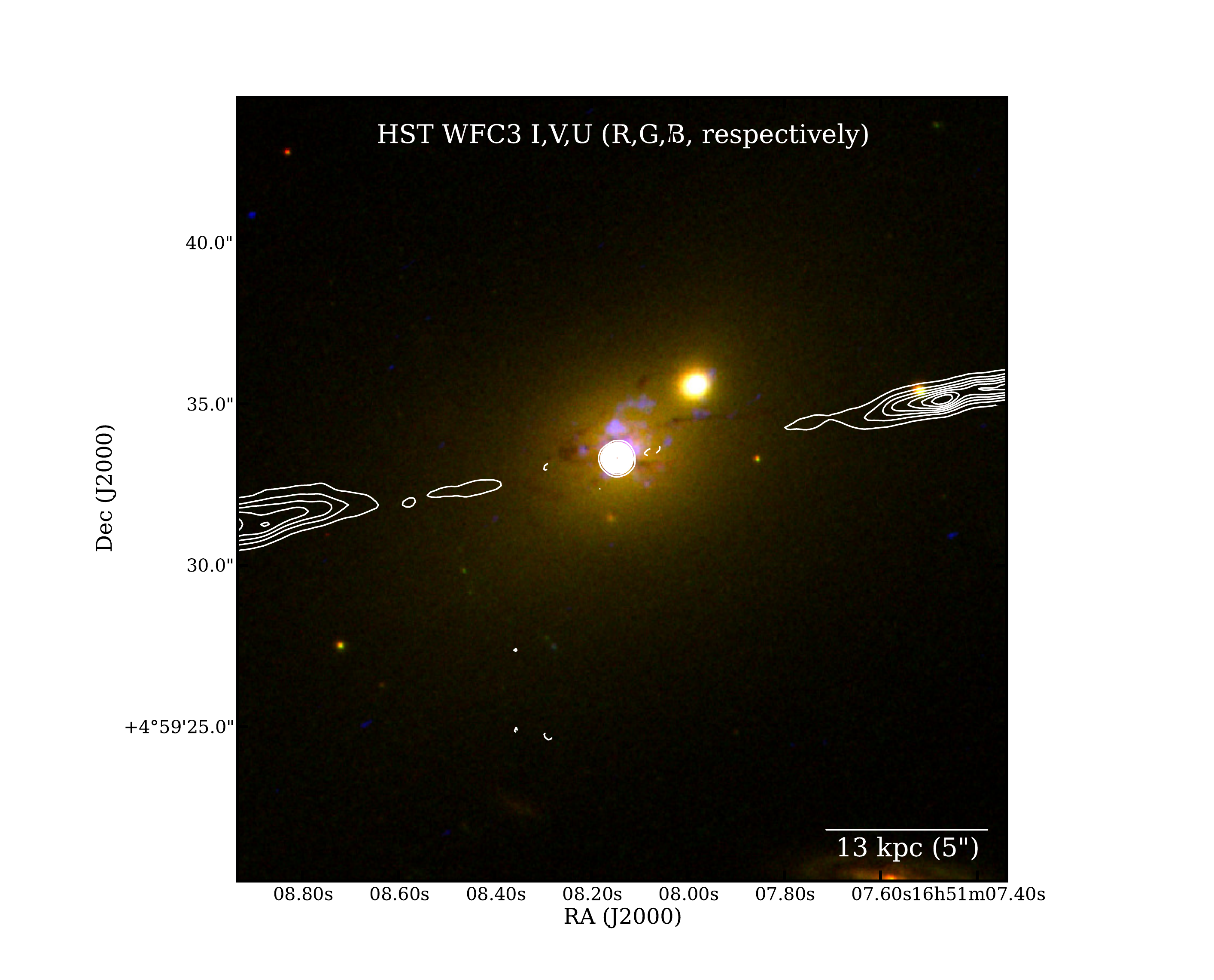}
\end{center}
\vspace*{-8mm}
\caption{Hercules A.  
 HST/WFC3/UVIS Red-Green-Blue image composed of the I, V, and U band images. 8.4 GHz JVLA radio contours are superposed in white. \label{fig_RGB}}
\end{figure*}

On the other hand, there is a secondary nucleus which in projection lies within the web of filaments (Figure~\ref{fig_V}) and  which may be experiencing a close encounter with or possibly merging with the host galaxy  and providing the gas/dust for the filaments.  Further, two of the western filaments extend out to near the position of the secondary nucleus. There is a blob of obscuration
just to the east of the secondary nucleus.  There is also a blue continuum feature which extends from the nucleus towards the secondary nucleus (Figs \ref{fig_U},\ref{fig_RGB}) which might be due to star formation in the acquired gas. The $I-V$ color map (Fig. \ref{fig_I-V}) shows that the
secondary nucleus and the host galaxy have similar colors. \citet{sadun93} suggested tentatively 
that the secondary nucleus is an S0 galaxy with an embedded disk and that the galaxy profile has been truncated. 
Thus, we favor the interpretation that the dusty filaments are due to gas acquired from an ongoing  merger/encounter with the secondary nucleus. 

None of the dusty filaments seem to lie in a preferred plane of the host galaxy. Combined with the "web-like" appearance 
of the filaments this suggests that  the cold gas has not had time to settle in the galaxy, or alternatively the gas has been
disturbed, e.g.,  by the hot ISM or the radio jets.

\subsection{Estimated Masses of Dust and Gas in the Filaments}

\begin{figure*}
\begin{center}
\includegraphics[scale=0.55]{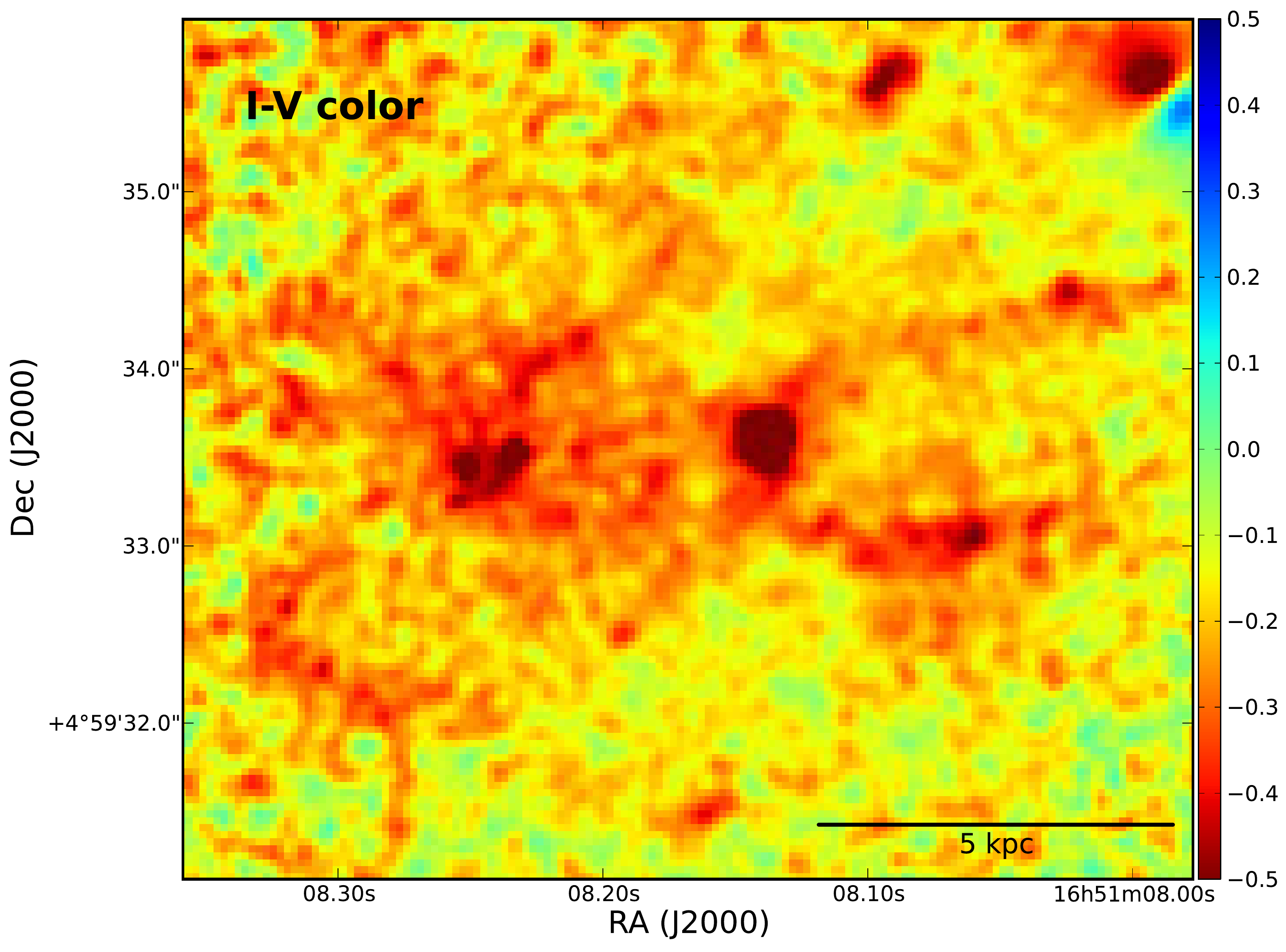}
\end{center}
\caption{Hercules A.   HST/WFC3/UVIS. Color image (I-V).  \label{fig_I-V}}
\end{figure*}

We use the $I-V$ color map (Fig. \ref{fig_I-V}) to roughly quantify the physical properties  of the gas and dust in the filaments. Following the methodology of \citet{sadler85} as adapted by \citet{dekoff00},  the lower-limit filamentary dust mass $M_\mathrm{dust}$ can be estimated by 

\begin{equation}
M_\mathrm{dust} = \Sigma \langle A_\lambda \rangle \Gamma^{-1}_\lambda,
\label{equation:dustmass}
\end{equation}

where $\Sigma$ is the area covered by the dusty filaments, $\langle A_\lambda \rangle$ is the  mean extinction due to dust within this area, and $\Gamma_\lambda$ is the  mass absorption coefficient, for which we adopt the Galactic value at $V$-band from \citet{vandokkum95};

\begin{equation}
\Gamma_V \approx 6 \times 10^{-6} \mathrm{mag~kpc}^2 M^{-1}_\odot .
\label{equation:massabsorption}
\end{equation}

A mean ``off-filament'' $I-V$ galaxy color is subtracted from a mean ``on-filament'' dust color,  yielding a color excess $E(I-V) \approx -0.4$ associated with the filaments that is (roughly) independent of  Milky Way and other internal extinction effects. After converting this color excess
to an extinction at $V$-band ($A_{V, \mathrm{dust}} \approx 0.775$ mag) following the Galactic $R_V=3.1$ law \citep{cardelli89}, 
equations \ref{equation:dustmass} and \ref{equation:massabsorption} are used to estimate a rough lower limit dust mass of $M_\mathrm{dust} \approx 3.8 \times 10^{6}~M_\odot$. If the filaments follow Galactic gas-to-dust ratios, the gas mass in the filaments would be roughly 100 times higher, or $\sim 4\times10^{8} M_\odot$. Finally, the extinction due to dust ($A_{V, \mathrm{dust}}$) can be used with the relation from \citet{predhel95} to estimate an associated neutral hydrogen column density of $\sim 1.4 \times 10^{21}$ cm$^{-2}$. 

\subsection{Star Formation \label{sec_SFR}}

\citet{nulsen05}  note that the high central electron density determined from Chandra observations gives a central cooling time consistent with cooling flow clusters. Additional details are provided by  Nulsen (2012, private communication) and \citet{tremblay13}. The cooling radius (where $t_{\rm cool} = 7.7$ Gyr) is at roughly 120 kpc (the number is not very well defined because the density profile is quite flat on the inside of the shock, so  $t_{\rm cool}$ varies slowly there).  These numbers imply a cooling power of about $10^{44}$ erg s\mone\ which gives  a cooling rate of 90 M$_\odot$ yr\mone\ (for kT = 4.5 keV).  However, the cooling times will change after the shock moves on, so the number is going to change on a timescale that is short compared to the cooling time. 

\begin{figure}
\begin{center}
\includegraphics[scale=0.37]{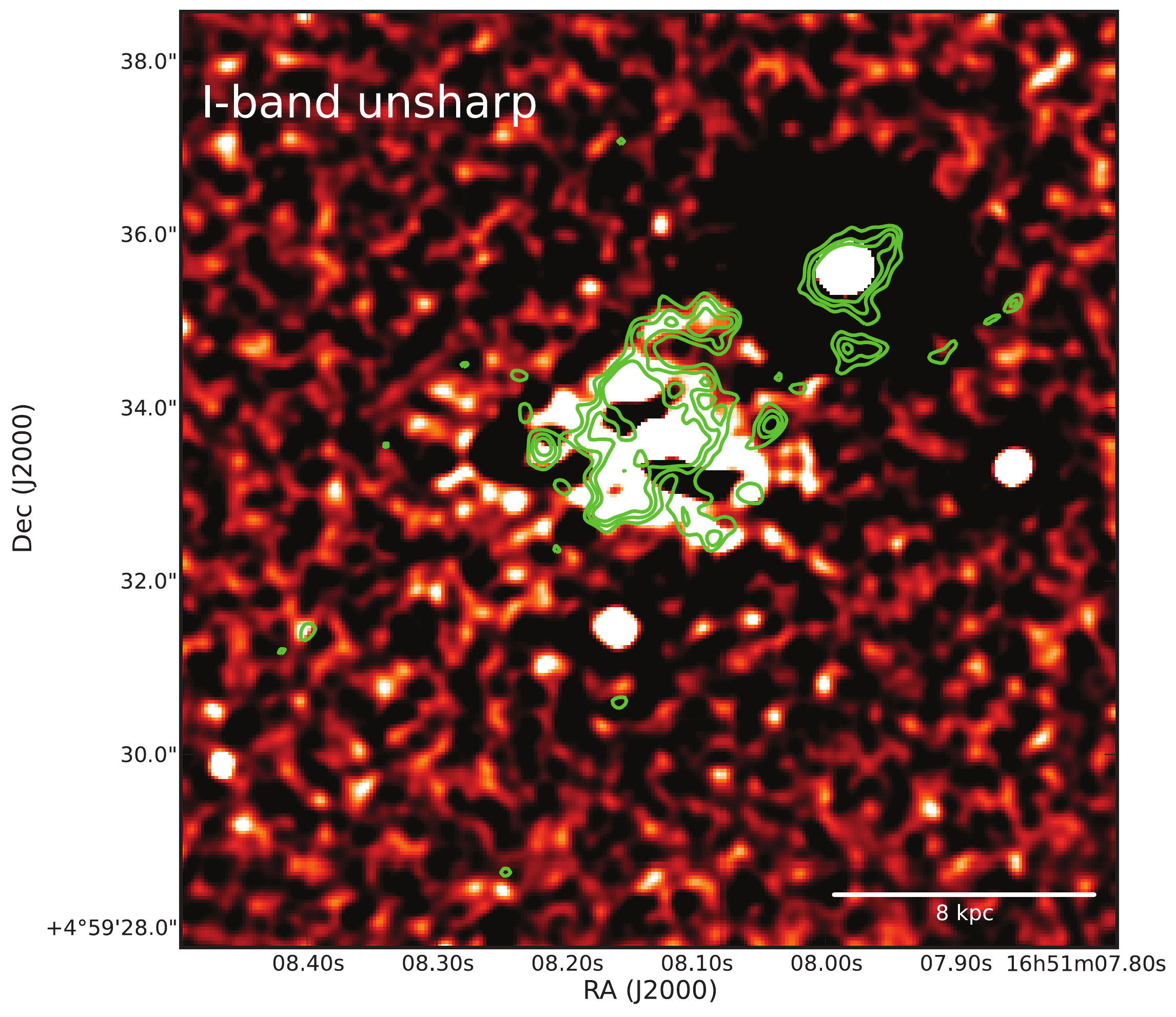}
\end{center}
\vspace*{-5mm}
\caption{Hercules A.  HST/WFC3/UVIS. I-band unsharp-masked image with contours of the U-band image superposed. \label{fig_I_unsharp}}
\end{figure}

The blue continuum light seen in Figures \ref{fig_U} and \ref{fig_RGB} may be due to low levels of star formation in the cold gas. High levels of star formation, i.e., a star burst, has already been ruled out \citep[e.g.,][]{dicken09}.   We can estimate star formation rates (SFRs) using  UV observations of photons from hot young stars, H$\alpha$ observations of gas photoionized by the young stars, and by Infrared emission from dust heated by the young stars \citep[e.g.,][]{kennicutt98, schmitt06}.  The H$\alpha$ luminosity is  L(H$\alpha$) $\simeq 1.9 \times 10^{41}$ erg s\mone\ \citep{buttiglione09}. Using the calibration of \citet{kennicutt98}  gives a SFR(H$\alpha$) $\sim 1.5 $ M$_\odot$ yr\mone. Because of the ``contamination" of the I-band filter with H$\alpha$+[NII] (\S \ref{sec_obs}, \ref{sec_nucleus}) there may be structure in the I-band image due to the emission line gas. In Figure \ref{fig_I_unsharp} we see faint structure in the I-band unsharp-masked image which is spatially coincident with the blue continuum seen in the U-band image. We expect the I-band continuum, extending from the old stellar component, to be smoothly distributed and mostly follow the smooth H-band surface brightness profile. Nevertheless, the I-band unsharp mask reveals clumpy and filamentary structures that we expect arise primarily from the H$\alpha$+[NII] contamination in the bandpass, which is not expected to be as smoothly distributed. It is therefore possible that the structure seen in Fig.  \ref{fig_I_unsharp} are H$\alpha$ filaments, which are cospatial with the U-band filament, consistent with H-alpha that has been excited by young stars. 
 The Spitzer 24 $\mu$m flux density is $2.0\pm0.2$ mJy \citep{dicken08}. Using the conversion of \citet{rieke09}  gives a star formation rate of SFR(24) $\simeq 4.1$ M$_\odot$ yr\mone. 
Using the U-band flux density corrected for galactic and internal extinction  (using the  Balmer decrement)  ($1.4 \times 10^{-16}$  erg sec\mone\ cm\mtwo\ \AA\mone) and converting to 
luminosity gives  L(U) $ \sim 3.3 \times 10^{21}$ W Hz\mone. Using the relation for U-band derived star formation rates from massive stars ($> 5$ M$_\odot$)  \citep{cowie97,hopkins98} gives  a star formation rate of $\sim 0.6$ M$_\odot$ yr\mone.
The values of SFR from the H$\alpha$, IR, and UV are consistent with star formation rates of  a few percent of the mass accretion rate derived from the cooling power as typically found in cluster cooling flows  \citep[e.g.,][]{mcnamara89, peterson06, odea08}.

\subsection{Radio and Optical Emission Line Properties \label{sec_line}}

The emission line nebula is faint and low excitation \citep{tadhunter93, buttiglione09, buttiglione10}.   \citet{buttiglione10} suggest that Hercules A is a member of a small class of Extremely Low  Excitation Galaxies (ELEG).  Radio powers as large as that  of Her A  (Log P$_{408} \simeq 28.3$ W Hz\mone) are generally only found in FRII radio galaxies. The H$\alpha$+[NII] emission line luminosity (Log $ L \simeq 34.6$ W), using the measurements of \citet{buttiglione09}  is about 1.5 orders of magnitude low for a FRII radio galaxy with that radio power. However, the emission line luminosity lies on an extrapolation to high radio power of the emission  line vs radio  relation for FRI radio galaxies found by \citet{baum95}.  Thus, the ratio of emission  line to radio power in Herc A is consistent with that of an FRI radio galaxy. 

Adopting a  1400 MHz core flux density from \citet{gizani03}  and a total flux density from \citet{kuhr81}  gives  the ratio of radio core to extended flux density $R \sim 8.5 \times 10^{-4}$ (log $R \simeq -3.1$). For comparison, sources with a similar high total radio power have a median log $R \simeq -2.4$ \citep{baum95}.  The value found in Hercules A is on the  low end of the distribution of R values. \citet{capetti11} 
 speculate that the low [OIII]/H$\beta$ combined with the low radio core to
extended radio emission ratio suggest that the nucleus of Hercules A has turned off. This is an interesting suggestion since 
intermittent radio activity could help to explain the unusual  radio properties of Hercules A, especially the series of shells/rings in the western lobe  \citep{gizani03}.

\citet{gizani02}  present EVN observations of the nucleus of Hercules A with 18 mas resolution which detect a compact radio source with a flux density of 14.6 mJy at 1.6 GHz.   In addition,  we detect the nucleus in our I-band image. The detection of the nucleus in the radio and
I-band suggests that the nucleus remains active though possibly at a much lower level than previously. 

Hercules A is in a  cooling flow cluster \citep{nulsen05}. The emission line nebulae in central dominant galaxies in cooling flows are almost always low excitation, and  [OIII]/H$\beta \sim 0.5$ is within the range of commonly observed values \citep[e.g.,][]{heckman89, crawford99, quillen08}.
  It seems probable  that the  emission line nebula in Herc A is  low excitation because of the lack of a strong ionizing AGN continuum and so the excitation is dominated by e.g., star formation, shocks, or processes occurring in the cooling flow environment.

However, radio galaxies which are fed by cold gas acquired in a merger tend to have radiatively efficient accretion and high excitation emission line nebulae \citep[e.g.,][]{baum95, hardcastle09,buttiglione10,best12}. The fact that the nebula in Hercules A is low excitation suggests that the dusty filaments are not providing significant amounts of fuel to the central supermassive  black hole. This might be because (1) the gas has not yet settled deep  into the galaxy nucleus, or (2) the cold gas in the nucleus is being removed, e.g., via entrainment in the radio jets.

\subsection{Timescales}\label{sec_time}

\citet{nulsen05} discovered a shock front in the hot ICM  with a radius of $\sim 160$ kpc. Based on fits to the jump in the X-ray  surface brightness profile they estimate a shock Mach number M $\simeq 1.65$. This gives an age for the outburst that created the shock of t$_s \sim 5.9 \times 10^7$ yr. 
If we assume a lobe expansion velocity of 0.03c \citep[e.g.,][]{alexander87,scheuer95,odea01,tremblay10}, the 200 kpc radio source radius implies a dynamical age of t$_d \sim2 \times 10^7$ yr.  For comparison, the spectral age of the electrons (due to radiative losses) at the end of the lobes is $\gae 1.3 \times 10^7$ yr \citep{gizani03} roughly consistent with the dynamical age estimate. The dynamical and radiative loss ages for the radio source are factors of $\sim 3$ and $\sim 4.5$ smaller than the shock age. This might be due to the uncertainties in the age estimates or it may suggest that there was a previous epoch of radio source activity $\sim 60$ Myr ago. 

 If the source has been recently reborn, the restarting jets would propagate faster through the evacuated lobe resulting in a shorter (model dependent) age of $1-4 \times 10^6$ yr \citep{gizani03}. If the dusty filaments are from the merger/interaction, they will need at least a dynamical time ($\gae 10^8$ yr) to reach the center.  The lack of a bright accretion disk suggests that the dusty filaments are not yet fueling the radio source and that the fueling is being done by the cooling flow (\S \ref{sec_line}). If this is the case, the stopping and restarting of the radio source would be associated with feedback and cooling timescales in the cooling flow.  The buoyant rise (75\% sound speed) and refilling times for both X-ray cavities are about $ 7 \times 10^7$ yr \citep{tremblay13} which is comparable to the age of the shock, but longer than the age of the current radio source. The difference in the ages of the X-ray cavities and the current radio source combined with the fact that the X-ray cavities are misaligned with the current  radio source by $\sim 65\deg$ would be consistent with the hypothesis that the cavities were created by the previous epoch of radio source activity (assuming that the cavities are created by the radio source and not by some other phenomena). However, that would suggest that the axis of the radio source changed by $\sim 65\deg$ in between the two epochs of activity. Such a large change in orientation may require a merger between supermassive black holes \citep[e.g.,][]{merritt02}. The black hole required for this merger may have been acquired in a merger prior to the current encounter/merger with the secondary nucleus. In addition, if the previous radio source is 60 Myr old, there should be some evidence for low frequency radio emission along the position angle of the X-ray cavities.  \citet{nulsen05} note  a buoyant bubble is formed in their model for the shock. If this mechanism would produce cavities misaligned with the radio source axis, then a swing in jet axis is not required. 

A possible scenario would be the following. About 60 Myr ago, a powerful radio source was generated propagating along a position angle of $35\deg$. This radio source created the shock and the cavities in the hot ICM.  A black hole merged with the central black hole changing the jet axis by $65\deg$ (in projection).  Either the source was ongoing and swung in position angle, or it had turned off and restarted at the new position angle of $ 100\deg$ about 20 Myr ago.  The source has maintained its position angle but the activity has been repetitive/unsteady on time scales of several Myr in order to  account for the structure in the lobes \citep{gizani03}.

\section{Constraints on Entrainment}

Our HST observations show dusty filaments aligned with the bases of the radio jets (e.g., Figures \ref{fig_V} and \ref{fig_dashed}).
Jets in FRI radio galaxies are launched with relativistic velocities but decelerate to sub-relativistic velocities by the time 
they reach  kpc scales \citep[e.g.,][]{bp84, odea85, laing93, urry95, giovannini01, kharb12}. It is widely thought that the deceleration is 
caused by entrainment of  thermal gas \citep[e.g.,][]{baan80, deyoung81, begelman82}. 
 The gas entrained could be either (1)  ambient gas entrained through a turbulent boundary layer \citep[e.g.,][]{bicknell84, bicknell94, deyoung86, deyoung96, komissarov90, perucho07, rossi08, wang09},
 (2) winds from stars within the volume  of the jet  \citep[e.g.,][]{komissarov94, bowman96}, or (3)  dense clouds which enter the jet \citep{fedorenko96,bosch12}. Laing and collaborators have modeled the collimation, surface brightness, and
polarization structure of  jets in FRIs in terms of  entraining and decelerating relativistic flows \citep{laing02a, laing02b, canvin04, canvin05, laing06}.
Previous evidence for entrainment has been suggested for 3C 277.3 \citep{vanbreugel85}  and Centaurus A (\citealt{graham81}; but cf \citealt{morganti91, graham98}).
 
The boundary layer develops at the interface between the jet and ambient medium and then spreads into both the 
ambient medium and the jet \citep{wang09}.  Using our HST observations we may be able to constrain how
 quickly the boundary layer develops, and how far it extends into the ambient medium. If the boundary
layer extends far into the ambient medium, this would suggest that there is a broader ``wind" which surrounds the jet and which
can influence the host galaxy. Here we consider the hypothesis that the dusty filaments are entrained by the radio jet 
and examine the implications. Because of the difference in morphology, we consider the two sides separately. Disclaimer: There could be a different physical relationship between the jet and filaments or none at all if the apparent spatial coincidence is due to chance projection. \citet{deyoung86}  finds that entrainment can occur in the radio source bow shock. Thus,  an alternative scenario could be that filaments were not entrained in a jet boundary layer, but were entrained in the bow shock of the radio source when it originally propagated through the host galaxy about 20 Myr ago (\S \ref{sec_time}). However, it is not clear that the dusty filaments would survive that long. 

\subsection{The Western Jet}

On the western side of the source, two dusty filaments align with the radio jet. The northern of the two filaments extends
about 4.7\arcsec\ from the nucleus, while the southern extends about 2\arcsec\ from the nucleus. 
 The radio jet does not become detectable until just beyond the extent of the western filaments.  This ``gap" is a common feature of FRI radio galaxies  \citep[e.g.,][]{bridle84, laing02a, laing02b, canvin04, canvin05, laing06}
 and in the Laing \& Bridle models is due to dimming caused by Doppler beaming of the radiation along the jet axis and thus out of our line-of-sight. 
We assume that there is an undetected jet which is continuous from the nucleus out to where the jet becomes detectable. 
 The continuity of the filament from the nucleus out to $\sim 4\arcsec$ suggests that the jet is entraining gas during the "gap" phase
of the jet (i.e., the inner 10 kpc). 

\begin{figure*}
\begin{center}
\includegraphics[scale=0.6]{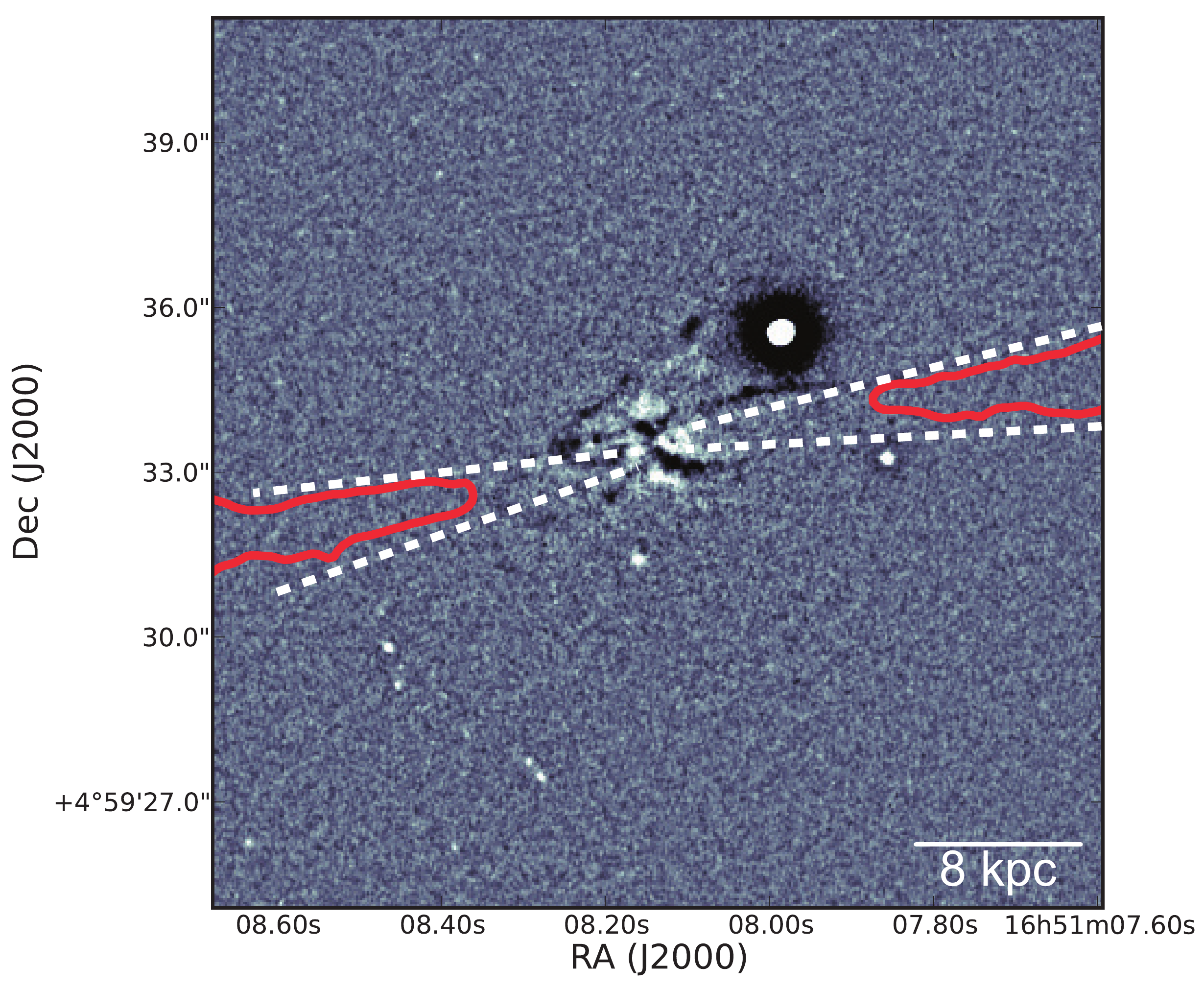}
\end{center}
\vspace*{-5mm}
\caption{Hercules A.    HST/WFC3/UVIS. V-band unsharp-masked image with low level radio contours of the JVLA image. Also shown in dashed lines is an extrapolation  inwards to the nucleus  of the radio jet to show roughly where the undetected inner jet might lie with respect to the dusty filaments. \label{fig_dashed}}
\end{figure*}

If the dusty gas is distributed in a cylindrical sheath, the higher column densities along our line of sight through the sides of the sheath will cause greater obscuration along the edges of the cylinder and thus the edges of the radio jets.  However, the asymmetry in the two filaments suggests that the dusty gas is not distributed homogeneously in  the sheath around the radio jet axis. 
In order to emphasize the spatial correlation between the filaments and radio jet we extrapolate the jet contours inwards towards the nucleus the radio jet contours (Figure ~\ref{fig_dashed}). We see that the filaments in projection lie along
the outside of the extrapolated radio jet. The separation between the two filaments (1\arcsec or 2.7 kpc)  is slightly larger than the width of the extrapolated jet. Jets on these scales in twin-jet sources tend to be expanding with distance from the nucleus \citep[e.g.,][]{bridle84, laing02a, laing02b, canvin04, canvin05, laing06}.
Thus, the "true"  inner jet at the location of the dusty filaments is likely to be narrower than the jet at 4\arcsec\  from the nucleus.
The dusty filaments are seen outside the visible extent of the jets at a distance of $\sim 1.3$ kpc from the jet axis. The width of the northern  filament is $\lae 0.1\arcsec$ (270 pc). The location of the observed dusty filaments relative to the jet axis depends on (1) where the gas is entrained, and (2) where it survives the process of entrainment and transport so that the filament is visible in absorption against the stellar background.  The lifetime of cold dusty  gas in high velocity flows is uncertain. Shocks which develop in such flows would heat/ionize  gas and destroy grains.  Thus, observational evidence for such cool gas is interesting. A possible relevant result is that  HI absorption in jet driven outflows in compact radio sources  reveals the presence of an atomic component to outflows with velocities up to $\sim 1000$ km s\mone\ \citep[e.g.,][]{holt08}. 
 The column density through the dusty filament determines its detectability in absorption against the stellar background. The filaments
can be made less detectable by mixing the cold gas and dust into a larger volume or by destroying the dust grains.  Dust grains will be destroyed by shocks with velocities greater than $\sim 100$ km s\mone\ \citep[e.g.,][]{jones96, welty02, jones11}. So, if the boundary layer is turbulent,  the dust may trace a low velocity ($v \lae 100$ km s\mone) outer edge in the boundary layer.   We don't see the dusty filaments widening or the filaments changing their projected distance from the axis of the jet as a function of distance from the nucleus as expected for a fully turbulent boundary layer. (Though there is a hint of a shortening  in
the distance of the northern filament from the jet axis around 4\arcsec\ from the nucleus.) Thus, in this early stage of entrainment, perhaps the entrainment is fairly laminar, or we preferentially see the dusty gas in regions of the flow which are laminar.

\subsection{The Eastern Jet}

Figures ~\ref{fig_V} and \ref{fig_dashed} show that unlike the relatively straight filaments which
bracket the jet on the western side, the eastern filaments are arc-like suggesting they are part of a bubble/shell or loop. Further the
eastern filaments are slightly misaligned from the jet axis (Figure~\ref{fig_dashed}). The inner filament is at PA $\sim 87\deg$ and the outer filament is at
PA $\sim 111\deg$. These arc-like structures resemble the bubbles seen in galactic super winds such as e.g., NGC 3079 \citep[e.g.,][]{cecil01} and NGC 6764 \citep[e.g.,][]{hota06}. However, it is not clear what the energy source would be for  winds in Hercules A. There is no bright accretion disk to drive a wind by radiation pressure (\S \ref{sec_line}), and the star formation rates are too small to drive a starburst-driven wind (\S \ref{sec_SFR}).  The only outflow that we know exists is the radio jet. On the other hand, if the structures we see on both sides are due to entrainment, then it is not clear why the morphologies on the two sides are so different.  

Although we don't understand the difference in morphology of the filaments on the two sides, for simplicity,  we hypothesize that the eastern filaments are also dragged out via entrainment in the outflowing radio jet. In this scenario, the dusty arcs are not pieces of shells blown out by a wind, but are instead loops of dusty filaments dragged out via entrainment. \citet{hatch06}  suggest that in the central cooling flow source NGC1275 (Perseus A), optical emission line filaments are dragged to  distances of order 20 kpc by buoyantly rising bubbles generated by the radio source. \citet{fabian08}  have suggested that the Perseus filaments are stabilized from disruption during their transport because they are threaded by magnetic fields. If this magnetic stabilization model is correct, it may also apply to the filaments in Herc A. The presence of blue continuum and H$\alpha$ emission associated with the filaments (\S \ref{sec_SFR}) suggests that the necessary free electrons will be available to freeze  the magnetic field to the filaments. If the filaments were stripped from the secondary nucleus, the linking of the magnetic field to the gas must have occurred in the secondary galaxy. 

 Occam's Razor favors a single explanation for the relation between the jets and dusty filaments on the  two sides of the source. However, because of the difference in morphology, it is possible that a different explanation is required for the two sides. One such scenario could be that only  the western filaments are entrained - producing their linear structure.  In this alternate scenario,  the eastern filaments are loops of dusty gas stripped from the secondary nucleus and  merely seen in projection against the jet. In this scenario the entrainment process is not required to produce two different filament morphologies. Though this would seem to require the secondary nucleus to pass through the plane containing the jets in order to produce a fortuitous alignment. 

As discussed in \S~\ref{sec_SFR} the unsharp-masked I band image (Figure \ref{fig_I_unsharp}) suggests that there is H$\alpha$ emission associated with the dusty filaments. If the filaments are indeed entrained in the jet boundary layer we might expect to see velocity structure perhaps along and/or across the filaments. In addition, there might be evidence for the ionization of the gas by  shocks due to turbulence or velocity sheer in the boundary layer.

For completeness, we note that if there are currents associated with the filaments and the radio jets \citep[e.g.,][]{benford78}, then Lorentz force type interactions might also be possible. 

\subsection{Implications for the Radio Morphology}

The differences in the structure of the radio jets on the two sides might be due to (1) differences in the way the outflows are generated by the supermassive black hole; or (2) differences in the way the outflows interact with their environments as they travel outward away from the galaxy.  It is not clear whether the mechanisms which produce the radio morphology are related to the mechanism which produce the morphology of the dusty filaments. If we assume that they are related, then it is possible to make the following argument. If the Hubble data showed that the morphology of the cold gas and dust mimicked that of the radio jet on the same side (i.e., dust shells on the side with radio shells, and straight filaments on the side with a collimated jet), that would be consistent with the idea that the differences in jet properties were established on small scales. However, the fact that the cold gas and dust has a different morphology than the radio emission on the same side supports the idea that the differences in radio properties are established on larger scales as the jets interact with their environments \citep[e.g.][]{gendre13}.

\section{SUMMARY}

We present U, V, and I band images of the host galaxy of Hercules A obtained with HST/WFC3/UVIS. We find a network of dusty filaments which are more complex and extended than seen in earlier HST observations. The filaments are associated with a faint blue continuum light (possibly from young stars) and faint H$\alpha$ emission.  Estimated star formation rates in the filaments are a few M$_\odot$ yr\mone\  which is a few percent of the  mass deposition rate  inferred from the Chandra-derived cooling power.  The total dust mass in the filaments is about $3.8 \times 10^6$ M$_\odot$. 

 It seems likely that the cold gas and dust has been stripped from a companion galaxy now seen as a secondary nucleus. There are dusty filaments aligned with the base of the jets on both eastern and western sides of the galaxy. The morphology of the filaments is different on the two sides - the western filaments are fairly straight, while the eastern filaments are mainly in two loop-like structures. We suggest that despite the difference in morphologies, both sets of filaments have been entrained in a slow moving boundary layer outside the relativistic flow. In the scenario, the entrainment has occurred during the ``gap" phase of jet evolution. As suggested by \citet{fabian08}  for NGC1275, magnetic fields in the filaments may stabilize them against disruption. 

We suggest a  speculative scenario to explain the relation of the radio source to the shock and cavities in the hot ICM.  About 60 Myr ago, a powerful radio source was generated propagating along a position angle of $35\deg$. This radio source created the shock and the cavities in the hot ICM.  A black hole merged with the central black hole changing the jet axis by $65\deg$ (in projection).  Either the source was ongoing and swung in position angle, or it had turned off and restarted at the new position angle of $ 100\deg$ about 20 Myr ago.  The source has maintained its position angle but the activity has been repetitive on time scales of several Myr in order to  account for the structure in the lobes \citep{gizani03}.

\acknowledgments

We are grateful to J. Stoke, M. Mutchler, Z. Levay, and L. Frattare for help with the HST 
proposal and Phase II file.  We thank the referee for helpful comments.  This work is based on observations made with the NASA/ESA Hubble Space Telescope, obtained  at the Space Telescope Science Institute, which is operated by the Association of Universities for Research in Astronomy, Inc., under NASA contract NAS 5-26555. These observations are associated with program \#13065. The National Radio Astronomy Observatory (NRAO) is operated by Associated Universities Inc. under cooperative agreement with the
National Science Foundation.

{\it Facilities:} \facility{JVLA}, \facility{HST (WFC3)}.

\end{document}